\documentclass[twocolumn,
superscriptaddress,
showpacs,preprintnumbers,
aps,
prb,
]{revtex4-1}

\usepackage{graphicx, color}
\usepackage{dcolumn}
\usepackage{bm}

\begin{document}


\title{Neutron-scattering study of yttrium iron garnet}

\author{Shin-ichi Shamoto}
\email[Corresponding author:]{shamoto.shinichi@jaea.go.jp}
\affiliation{Advanced Science Research Center, Japan Atomic Energy Agency (JAEA), Tokai, Naka, Ibaraki 319-1195, Japan}

\author{Takashi U. Ito}
\affiliation{Advanced Science Research Center, Japan Atomic Energy Agency (JAEA), Tokai, Naka, Ibaraki 319-1195, Japan}

\author{Hiroaki Onishi}
\affiliation{Advanced Science Research Center, Japan Atomic Energy Agency (JAEA), Tokai, Naka, Ibaraki 319-1195, Japan}

\author{Hiroki Yamauchi}
\affiliation{Materials Sciences Research Center, Japan Atomic Energy Agency (JAEA), Tokai, Naka, Ibaraki 319-1195, Japan}

\author{Yasuhiro Inamura}
\affiliation{J-PARC Center, Japan Atomic Energy Agency (JAEA), Tokai, Ibaraki 319-1195, Japan}

\author{Masato Matsuura}
\affiliation{Neutron Science and Technology Center, Comprehensive Research Organization for Science and Society (CROSS), Tokai, Naka, Ibaraki 319-1106, Japan}

\author{Mitsuhiro Akatsu}
\affiliation{Dept. of Phys, Niigata Univ., Niigata, Niigata 950-2181, Japan}


\author{Katsuaki Kodama}
\affiliation{Materials Sciences Research Center, Japan Atomic Energy Agency (JAEA), Tokai, Naka, Ibaraki 319-1195, Japan}

\author{Akiko Nakao}
\affiliation{Neutron Science and Technology Center, Comprehensive Research Organization for Science and Society (CROSS), Tokai, Naka, Ibaraki 319-1106, Japan}

\author{Taketo Moyoshi}
\affiliation{Neutron Science and Technology Center, Comprehensive Research Organization for Science and Society (CROSS), Tokai, Naka, Ibaraki 319-1106, Japan}

\author{Koji Munakata} 
\affiliation{Neutron Science and Technology Center, Comprehensive Research Organization for Science and Society (CROSS), Tokai, Naka, Ibaraki 319-1106, Japan}

\author{Takashi Ohhara}
\affiliation{J-PARC Center, Japan Atomic Energy Agency (JAEA), Tokai, Ibaraki 319-1195, Japan}

\author{Mitsutaka Nakamura}
\affiliation{J-PARC Center, Japan Atomic Energy Agency (JAEA), Tokai, Ibaraki 319-1195, Japan}

\author{Seiko Ohira-Kawamura}
\affiliation{J-PARC Center, Japan Atomic Energy Agency (JAEA), Tokai, Ibaraki 319-1195, Japan}

\author{Yuichi Nemoto}
\affiliation{Grad. Sch. of Sci. Tech. Niigata Univ., Niigata, Niigata 950-2181, Japan}

\author{Kaoru Shibata}
\affiliation{J-PARC Center, Japan Atomic Energy Agency (JAEA), Tokai, Ibaraki 319-1195, Japan}

\date{\today}

\begin{abstract}
The nuclear and magnetic structure and full magnon dispersions of yttrium iron garnet Y$_3$Fe$_5$O$_{12}$ have been studied using neutron scattering. The refined nuclear structure is distorted to a trigonal space group of $R\bar{3}$. The highest-energy dispersion extends up to 86 meV. The observed dispersions are reproduced by a simple model with three nearest-neighbor-exchange integrals between 16$a$ (octahedral) and 24$d$ (tetrahedral) sites, $J_{aa}$, $J_{ad}$, and $J_{dd}$, which are estimated to be 0.00$\pm$0.05, $-2.90$$\pm$0.07, and $-0.35$$\pm$0.08 meV, respectively. The lowest-energy dispersion below 14 meV exhibits a quadratic dispersion as expected from ferromagnetic magnons. The imaginary part of $q$-integrated dynamical spin susceptibility $\chi$"($E$) exhibits a square-root energy-dependence at low energies. The magnon density of state is estimated from the $\chi$"($E$) obtained on an absolute scale. The value is consistent with the single chirality mode for the magnon branch expected theoretically. 
\end{abstract}

\maketitle

\section{INTRODUCTION} 

Yttrium iron garnet (YIG) with a chemical composition of Y$_3$Fe$_5$O$_{12}$ is a well-known ferrimagnetic insulator for various applications, recently expanding to spintronic devices \cite{Kajiwara, Chumak}. The spin current is excited as a flow of spin-angular-momentum of magnon thermally depending on the magnon dispersion. The spin current produces a voltage on an attached electrode such as platinum via the inverse spin Hall effect \cite{Heremans}. This phenomenon is called the spin Seebeck effect \cite{Uchida}. Especially, a sample configuration of the ferromagnetic material and the electrode with a thermal gradient along a longitudinal direction is called a longitudinal spin Seebeck effect (LSSE) \cite{Kikkawa}. A recent detailed study of LSSE on YIG showed that the magnetic field dependence has good agreement with that expected from bulk-YIG magnon dispersion\cite{Kikkawa, Kikkawa16}. The magnetic field produces a gap in the ferromagnetic dispersion, resulting in the reduction of thermally excited spin current. The theoretical model is based on basic magnon parameters of YIG, which are the quadratic magnon dispersion and the magnon density of states (MDOS, ${\mathcal D}_M$). The magnon dispersons have been measured by inelastic neutron scattering (INS) \cite{Plant, Plant2}. Theoretical studies \cite{Ohnuma, Barker} suggest that the splitting of two types of modes $\chi"_{xy}$ and $\chi"_{yx}$ plays an important role for the LSSE. They correspond to negative and positive chirality (polarization) modes, respectively \cite{Barker}. In a sub-unit cell with five Fe spins of this ferrimagnet as shown in Fig. 1, there are three up spins and two down spins, corresponding to the three positive and two negative chirality modes, respectively. One of the mode carries one directional spin current, whereas the other does the opposite direction. Therefore, mixing of two modes cancels the spin current, resulting in a reduction of the spin Seebeck effect.  This mechanism is theoretically proposed to play an important role for the degradation of LSSE at high temperatures \cite{Barker}.

\begin{figure}[b] 
\includegraphics[width=6cm,clip]{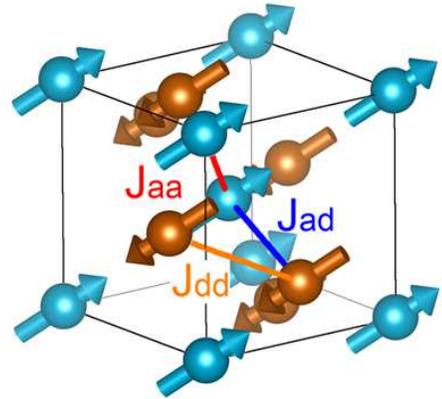}
\caption{(Color online) Fe spins in a sub-unit cell, which is 1/8 of a cubic unit cell ($Ia\bar{3}d$) with 40 Fe spins. Blue and brown arrows are spins at 16$a$ (octahedral) and 24$d$ (tetrahedral) sites for $Ia\bar{3}d$, respectively. Three nearest-neighbor-exchange integrals between 16$a$ and 24$d$ sites, $J_{aa}$, $J_{ad}$, and $J_{dd}$, are shown by red, blue, and orange lines, respectively.}
\label{fig:1}
\end{figure}

As for the basic properties of YIG, there have been a plenty of reports \cite{Cherepanov}. The nuclear and magnetic structure, however, has been studied so far only by using a powder sample \cite{Rodic}. According to a study under a magnetic field, the nuclear structure is distorted from a cubic to a trigonal space group. Under a magnetic field, a single crystal measurement is suitable to remove the magnetic domain walls homogeneously for the precise nuclear and magnetic structure determination. In addition, even the magnon dispersions have never been studied above 55 meV due to a technological limitation in INS measurements \cite{Plant2}. The exchange integrals of YIG have been estimated under the limitation \cite{Plant, Plant2, Cherepanov}. Current high-efficiency INS spectrometers at pulsed-neutron sources enable us to access high energy $E$ above 55 meV even with a small crystal. According to the theory of YIG magnons \cite{Barker}, the mode mixing for the lowest-$E$ branch becomes important for LSSE. For an anti-ferromagnet, the lowest-$E$ dispersion is known to have doubly degenerated modes. On the other hand, the lowest-$E$ dispersion of YIG is theoretically predicted to have only single mode of positive chirality \cite{Barker}. It is challenging for us to check the mode number from the MDOS on an absolute scale, due to various factors such as neutron absorption in a sample. Because of the difficulties, there has been no report of absolute MDOS for YIG. 

Here, we study all these important issues by using neutron scattering. As for the MDOS estimation, we introduce an approximated dynamical structure factor and effective reciprocal space volume to simplify the absolute estimation in addition to the numerical calculation of absorption coefficients. The INS probability on a magnet can be expressed by Fermi's golden rule, which includes the MDOS of the final states. This is the same as the phonon density of states for the phonon case, which is often revealed by INS. Based on this method, the MDOS of YIG was estimated from the observed scattering intensity. So far, the sum rule of the $q$-integrated scattering function $S(E)$ after energy integration is well known to be proportional to $S(S+1)$\cite{Shamoto}. Therefore, the $q$-integrated dynamical spin susceptibility $\chi$"($E$) normalized by $g^2{\rm \mu^2_B}S(S+1)$ becomes the MDOS at $T$ = 0 K, where the energy integration results in unity. 
To check the validity of our estimation, the simple quadratic dispersion model\cite{Kikkawa} is used to estimate the MDOS. In the low energies below 14 meV, the magnon dispersion is well described by a simple quadratic function of wave vector. The observed lowest-$E$ dispersion is fitted by the simple quadratic dispersion with a stiffness constant $D$. In addition, $D$ is also checked by the exchange integrals obtained from the whole magnon spectrum in our experiment. Then the relation between the MDOS and $\chi$"($E$) is discussed quantitatively based on the stiffness constant $D$.

\section{EXPERIMENTAL PROCEDURES}

Single crystals of YIG were grown by a traveling solvent floating zone furnace\cite{Kimura} with four halogen lamps (FZ-T-4000-H-II-S-TS, Crystal Systems Co., Ltd.) at a rate of 0.6-1.0 mm/h under air-flow of 2 L/min.  
The typical crystal size was 5.5 mm diameter and 47 mm length with a weight of 5.8 g. The mosaic spreads of both of the crystal ends were about 1.5 degrees based on x-ray Laue measurement. 

All magnon excitations of YIG in a wide $E$ range from 0.05 to 86 meV were observed by INS measurements with three different types of time-of-flight spectrometers, 4D-Space Access Neutron Spectromete (4SEASONS)\cite{Kajimoto}, Cold-neutron disk-chopper spectrometer (AMATERAS)\cite{Nakajima}, Biomolecular Dynamics Spectrometer (DNA)\cite{Shibata} at Japan Proton Accelerator Research Complex (J-PARC) Materials Life Science Experimental Facility. Based on the magnon excitations, the nearest-neighbor-exchange integrals were estimated by using {\sc spinw} software \cite{spinw} based on the linear spin wave theory with Holstein-Primakoff approximation. The observed scattering patterns were simulated by {\sc Horace} \cite{horace} and {\sc spinw} softwares. 
On the other hand, the magnetic Bragg peak intensities were measured by Extreme Environment Single Crystal Neutron Diffractometer (SENJU)\cite{SENJU}. The collected data were processed with the software {\sc stargazer}\cite{STARGazer}. Their intensities were analyzed by {\sc fullprof}\cite{FullProf}. The nuclear and magnetic structures were generated by {\sc vesta}\cite{VESTA}. Errors in the magnetic structure section are shown in the parentheses by the corresponding digits.   

Here, we define the scattering wave vector ${\bf Q}$ as ${\bf Q}={\bf q}+{\bf G}$, where ${\bf Q}$=$Q_a(2,-1,-1)+Q_b(1,1,1)+Q_c(0,-1,1)$, ${\bf q}$=$q_a(2,-1,-1)+q_b(1,1,1)+q_c(0,-1,1)$is defined in the crystal setting Brillouin zone, and ${\bf G}$ is a reciprocal lattice vector such as (220) for a cubic unit cell of $Ia\bar{3}d$ with a lattice parameter $a$=12.36 \AA. $Q_a$, $Q_b$, $Q_c$, $q_a$, $q_b$, and $q_c$ are in reciprocal lattice units (r.l.u.). The horizontal scattering plane was set in the ($Q_a$,$Q_b$,0) zone. Note that this crystal-setting Brillouin zone with 1 r.l.u.$^3$ is 6 times larger than the original Brillouin zone. Incident energies $E_i$ of the multi-$E_i$ mode\cite{Nakamura} were 12.5, 21.5, 45.3, and 150.0 meV at 4SEASONS, whereas it was 15.0 meV at AMATERAS. The incoherent $E$ widths of FWHM at $E$=0 meV were 0.67$\pm$0.04, 1.30$\pm$0.06, 3.9$\pm$0.1, 20.0$\pm$0.2, and 1.07$\pm$0.02 meV, respectively. As for the INS measurement at DNA, the horizontal scattering plane was set in the (0,$Q_b$,$Q_c$) zone. The final energy $E_f$ of DNA was 2.08 meV. The incoherent $E$ width of FWHM was 3.44$\pm$0.02 $\mu$eV at $E$=0 meV ($Q$=1.44 \AA$^{-1}$). The INS measurements were carried out at $T \approx$ 20 K. It typically took about 1 day for one measurement under a proton beam power of about 300 kW at J-PARC. 
The neutron absorption coefficients $A^*$ of our YIG single crystal were estimated based on our numerical calculation, and were typically 0.67 and 0.54 for 4SEASONS/AMATERAS and DNA, respectively. 

\section{RESULTS AND DISCUSSION}
\subsubsection{Nuclear and magnetic structure}

The nuclear structure of YIG is reported to distort from cubic to trigonal symmetries under a magnetic field along [111]$_{cubic}$ \cite{Rodic}. In order to check the trigonal distortion, nuclear and magnetic structure of YIG was studied using a single crystal at about 295 K under a magnetic field ($B$ $\approx $ 0.1 T) along [111]$_{cubic}$ with a pair of permanent magnets to remove the magnetic domain walls. The magnetic field at the sample position was measured by a Hall effect sensor. Intensities of 727 reflections (with the conditions $I > 3 \sigma I$ and $\sin \theta/\lambda < 0.85$ \AA) were well refined with a trigonal space group ($R\bar{3}$, No. 148: hexagonal setting) \cite{Rodic} with lattice parameters $a$=17.50227(55) \AA \ and $c$=10.73395(29) \AA. Observed nuclear and magnetic Bragg peak intensities are shown in Fig. 2 as a function of calculated intensities. All the magnetic moments align along the [001]$_{hexagonal}$ ([111]$_{cubic}$) direction parallel to $B$. The obtained nuclear and magnetic structure is shown in Fig. 3. The refined crystallographic parameters with reliability factors $R_{F^2}$ = 9.85\% and $R_F$ = 7.07\% are listed in Table 1. They were almost consistent with the reported values \cite{Rodic}. Note that the occupancy $g$ of O7$_{18f}$ was fixed because the $g$ value exceeded 1.00 slightly within the error during the refinement. The chemical composition of the present YIG crystal was Y$_{2.84(9)}$Fe$_{5}$O$_{11.57(21)}$. The deficiency of the Y$^{3+}$ ion is almost compensated by the oxygen deficiency for the Fe valence of +3 within the error. The obtained magnetic moments were 3.50 ${\rm \mu_B}$$\pm$0.17 ${\rm \mu_B}$ and 3.37 ${\rm \mu_B}$$\pm$0.17 ${\rm \mu_B}$ at octahedral and tetrahedral sites, respectively, where ${\rm \mu_B}$ is the Bohr magneton. The total magnetization of 3.1(6) ${\rm \mu_B}$/f.u. is consistent with the magnetization 3.05 ${\rm \mu_B}$/f.u. under $B$ = 1 T, although the obtained magnetic moments under $B$ $\approx$ 0.1 T are smaller than 4.47 ${\rm \mu_B}$$\pm$0.04 ${\rm \mu_B}$ and 4.02 ${\rm \mu_B}$$\pm$0.05 ${\rm \mu_B}$ \cite{Rodic}. These discrepancies can be attributed to the remaining magnetic domain walls of the powder sample in the previous study \cite{Rodic}. The slightly larger trigonal lattice distortions observed here compared with the previous ones reduce the observed magnetic moments in the present analysis due to the overlapping of Bragg peaks. In this sense, it is very important to determine the nuclear structure of YIG precisely for the estimation of the magnetic moments. Although the atomic distortions from cubic to trigonal symmetries are observed, the following magnon dispersions are discussed in the cubic symmetry of $Ia\bar{3}d$ for simplicity.   

\begin{figure}[h] 
\includegraphics[width=8cm,clip]{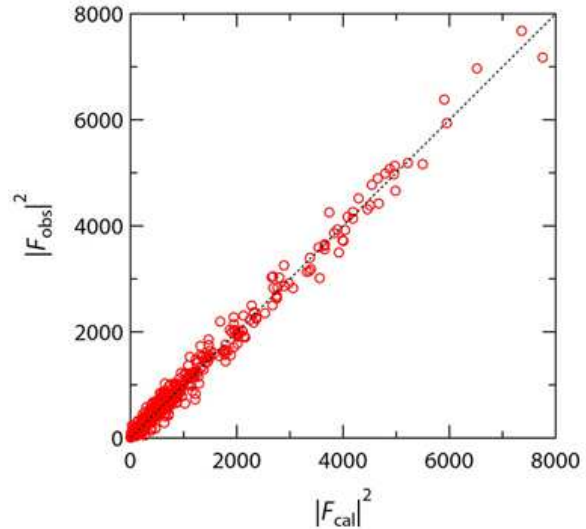}
\caption{(Color online) Observed nuclear and magnetic Bragg peak intensities as a function of calculated intensities with a trigonal space group ($R\bar{3}$).}
\label{fig:2}
\end{figure}
 
\begin{table*}[hbtp]
 \caption{The parameters of the nuclear and magnetic structure of YIG at about 295 K under $B$ $\approx$ 0.1 T in the space group $R\bar{3}$. Errors are shown in parentheses by the corresponding digits. The occupancies $g$ of Fe and O7$_{18f}$ were fixed as indicated by ``fix''.} 
 \label{table1}
 \centering
\begin{ruledtabular}
 \begin{tabular}{lrrrrrrr}
Atoms & \multicolumn{3}{c}{Fractional coordinates}&$B$ (\AA$^2$)&$g$&$\mu_z$ (units of ${\rm \mu_B}$)\\
&\multicolumn{1}{c}{$x$} & \multicolumn{1}{c}{$y$} & \multicolumn{1}{c}{$z$}\\
\hline
Y1$_{18f}$ & 0.1255 (3) & 0.0005 (3) & 0.2497 (2) & 0.245 (53) & 0.943(22) & \\
Y2$_{18f}$ & 0.2911 (3) & 0.3333 (3) & 0.5834 (2) & 0.245 (53) & 0.948(22) & \\
Fe$_{3a}$ & 0 & 0 & 0 & 0.243 (27) & 1.00(fix) & 3.50(17) \\
Fe$_{3b}$ & 0 & 0 & 0.5 & 0.243 (27) & 1.00(fix) & 3.50(17) \\
Fe$_{9d}$ & 0 & 0.5 & 0.5 & 0.243 (27) & 1.00(fix) & 3.50(17) \\
Fe$_{9e}$ & 0.5 & 0 & 0 & 0.243 (27) & 1.00(fix) & 3.50(17) \\
Fe1$_{18f}$ & 0.2084(2) & 0.1672(2) & 0.4166(2) & 0.315 (28) & 1.00(fix) & $-$3.37(17) \\
Fe2$_{18f}$ & 0.2912(2) & -0.1670(2) & 0.5832(2) & 0.315 (28) & 1.00(fix) & $-$3.37(17) \\
O1$_{18f}$ & 0.0877 (3) & 0.0920 (4) & 0.1210 (2) & 0.344 (28) & 0.993(18) &  \\
O2$_{18f}$ & 0.2622 (4) & 0.1158 (4) & 0.3230 (3) & 0.344 (28)  & 0.941(17) & \\
O3$_{18f}$ & $-$0.4212 (3) & $-$0.3721 (4) & 0.5444 (2) & 0.344 (28)  & 0.991(20) & \\
O4$_{18f}$ & 0.4867 (4) & 0.0953(4) & 0.4188 (3) & 0.344 (28)  & 0.940(17) & \\
O5$_{18f}$ & $-$0.0042 (4) & $-$0.0904 (4) & 0.3798 (3) & 0.344 (28)  & 0.963(18) & \\
O6$_{18f}$ & 0.1453 (4) & $-$0.1154 (4) & 0.1773 (3) & 0.344 (28)  & 0.940(18) & \\
O7$_{18f}$ & $-$0.0490 (3) & 0.3717 (4) & $-$0.0451 (2) & 0.344 (28)  & 1.00(fix) & \\
O8$_{18f}$ & 0.3899 (4) & $-$0.0967 (4) & 0.0809 (3) & 0.344 (28)  & 0.933(17) & \\
\end{tabular}
\end{ruledtabular}
\end{table*}

\begin{figure}[h] 
\includegraphics[width=8cm,clip]{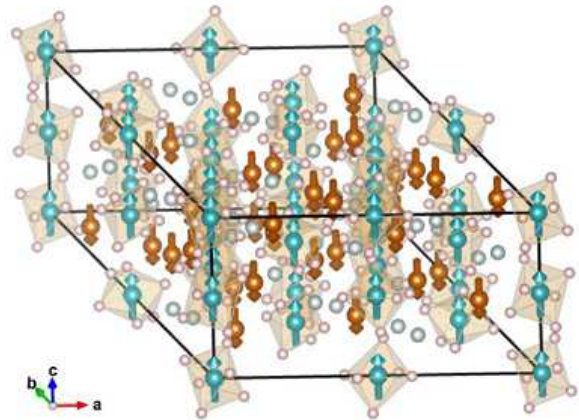}
\caption{(Color online) Obtained nuclear and magnetic structure of YIG in a trigonal unit cell of $R\bar{3}$. Blue and brown arrows are iron spins for octahedral and tetrahedral sites, respectively, small pink spheres are oxygen, and pale blue spheres are yttrium.}
\label{fig:3}
\end{figure}
\subsubsection{Magnon dispersions}

A magnon excitation in YIG is observed at ${\bf q}$ in a reciprocal space deviating from the $\Gamma$ point at a finite energy transfer $E$. It forms a three-dimensional (3D) $q$ spherical shell at $E$ due to the 3D interactions of localized spins as shown in Fig. 4, where the measured magnetic excitations were extracted as a two-dimensional slice from the INS data set. 
The normalized intensities $CI_{obs}({\bf Q},E)$ at around (220) thinly sliced along $Q_a$, $Q_b$, and $Q_c$ at $E$ $\approx$ 12 meV are shown in Fig. 5. The averaged integrated intensities $CI_{obs}({\bf Q}, E)$ ($\Delta V_Q$= 0.01) along $Q_a$, $Q_b$, and $Q_c$ were 0.87$\pm$0.12, 1.40$\pm$0.14, and 1.92$\pm$0.20, respectively. The peak positions from (220) along $q_a$, $q_b$, and $q_c$ were 0.35$\pm$0.01, 0.343$\pm$0.003, and 0.332$\pm$0.004 \AA$^{-1}$, respectively. These positions suggest that the magnon dispersion is nearly isotropic. 
  
The $q$ positions of magnetic excitations were determined by the fittings of Gaussian functions for thinly sliced $Q$ scans, as shown in Fig. 5. On the other hand, $q$-integrated magnon intensity was obtained through the integration of one 3D spherical-shell excitation. See the Appendix for the details of our absolute intensity estimation.

\begin{figure}[t] 
\includegraphics[width=8cm,clip]{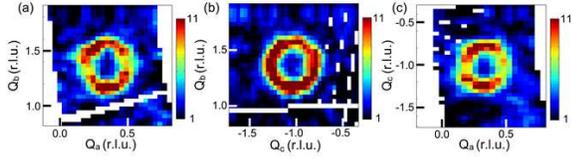}
\caption{(Color online) One of the magnon excitations of YIG at (220) ($(Q_a, Q_b, Q_c)$=(1/3, 4/3, $-$1)) in the $E$ range from 5 to 20 meV measured by $E_i$=45.3 meV. (a) $Q_a$-$Q_b$ contour map of $-1.05<Q_c<-0.95$. (b) $Q_c$-$Q_b$ contour map of $0.3<Q_a<0.4$. (c) $Q_a$-$Q_c$ contour map of $1.3<Q_b<1.4$. White areas are no detector regions. The color bars are in units of mbarn sr$^{-1}$meV$^{-1}$r.l.u.$^{-3}$. }
\label{fig:4}
\end{figure}
\begin{figure}[t] 
\includegraphics[width=8cm,clip]{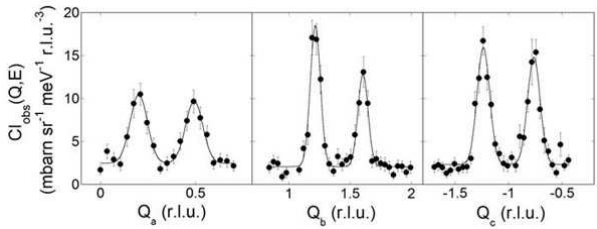}
\caption{Normalized intensity $CI_{obs}({\bf Q},E)$ in Fig. 4 as a function of ${\bf Q}$ in the $E$ range from 5 to 20 meV. (a) $Q_a$ scan of $1.3< Q_b< 1.4$ and  $-1.05< Q_c<-0.95$. (b) $Q_b$ scan of $0.3< Q_a<0.4$ and  $-1.05< Q_c<-0.95$. (c) $Q_c$ scan of $0.3< Q_a<0.4$ and $1.3< Q_b< 1.4$.}
\label{fig:5}
\end{figure}
\begin{figure}[t] 
\includegraphics[width=8cm,clip]{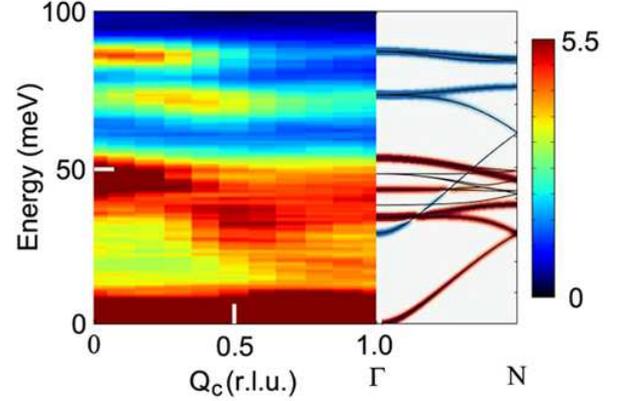}
\caption{(Color online) Wide $E$ range magnon dispersions in $Q_c$-$E$ space. Left: observed pattern as a function of $Q_c$ measured by $E_i$=150.0 meV in the range of  $-0.5< Q_a<3$ and $1< Q_b< 3$. Right: magnon dispersions along the same direction calculated from $\Gamma$ to $N$ at (123) by {\sc spinw} with the three nearest-neighbor-exchange integrals estimated here. The brown (blue) coloring denotes the positive (negative) chirality mode. The color bar is in units of mbarn sr$^{-1}$meV$^{-1}$r.l.u.$^{-3}$.}
\label{fig:6} 
\end{figure}

\begin{figure*}[tbp] 
\includegraphics[width=14cm,clip]{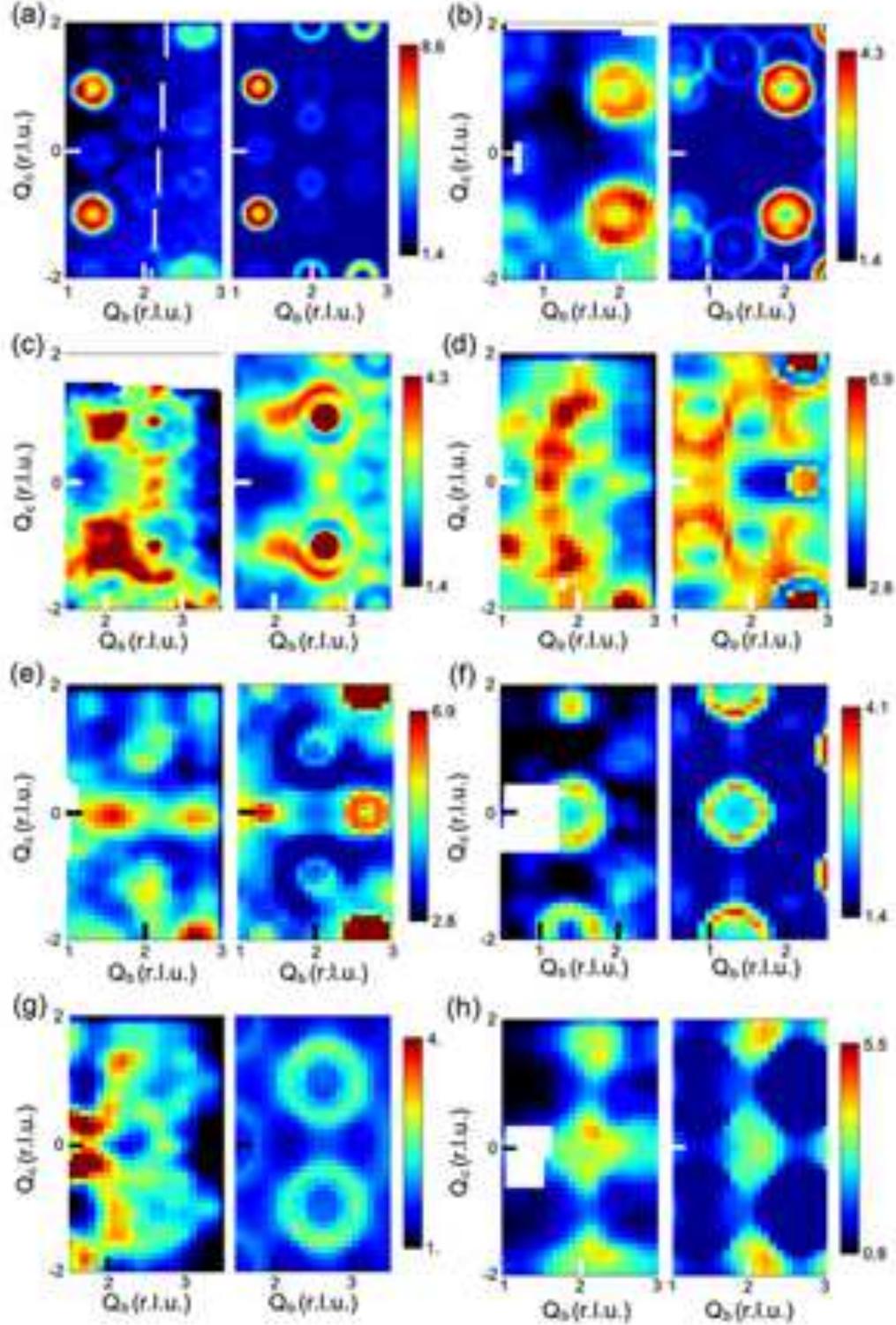}
\caption{(Color online) Constant-$E$ cuts of magnon spectra in the $Q_b$-$Q_c$ plane with an $E$ width of 10 meV. Left and right plots show observed and simulated patterns, respectively. The transfer energies are (a) 12, (b) 25, and (c) 33 meV for $E_i$=45.3 meV, and (d) 40, (e) 50, (f) 60, (g) 70, and (h) 85 meV for $E_i$=150.0 meV. The corresponding $h$ values in (2$h$, $-h$, $-h$) are about 0.2, 0.9, 1.7, 0.65, 0.95, 1.25, 1.6, and 1.9, respectively. The color bars are for observed patterns in units of mbarn sr$^{-1}$meV$^{-1}$r.l.u.$^{-3}$.}
\label{fig:7} 
\end{figure*}

\begin{figure}[h] 
\includegraphics[width=8cm,clip]{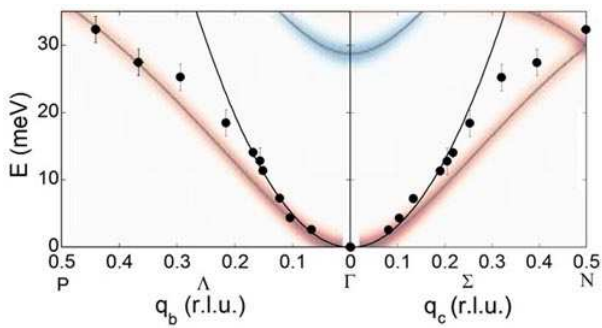}
\caption{(Color online) The lowest-$E$ magnon dispersion along the $\Lambda$ and $\Sigma$ directions. The solid line is the fitting with Eq. (1). The calculated dispersions with exchange integrals are also shown by pale blurry lines in the same $Q$-$E$ space. Brown (blue) denotes the positive (negative) chirality mode.}
\label{fig:8}
\end{figure}

\begin{figure}[h] 
\includegraphics[width=8cm,clip]{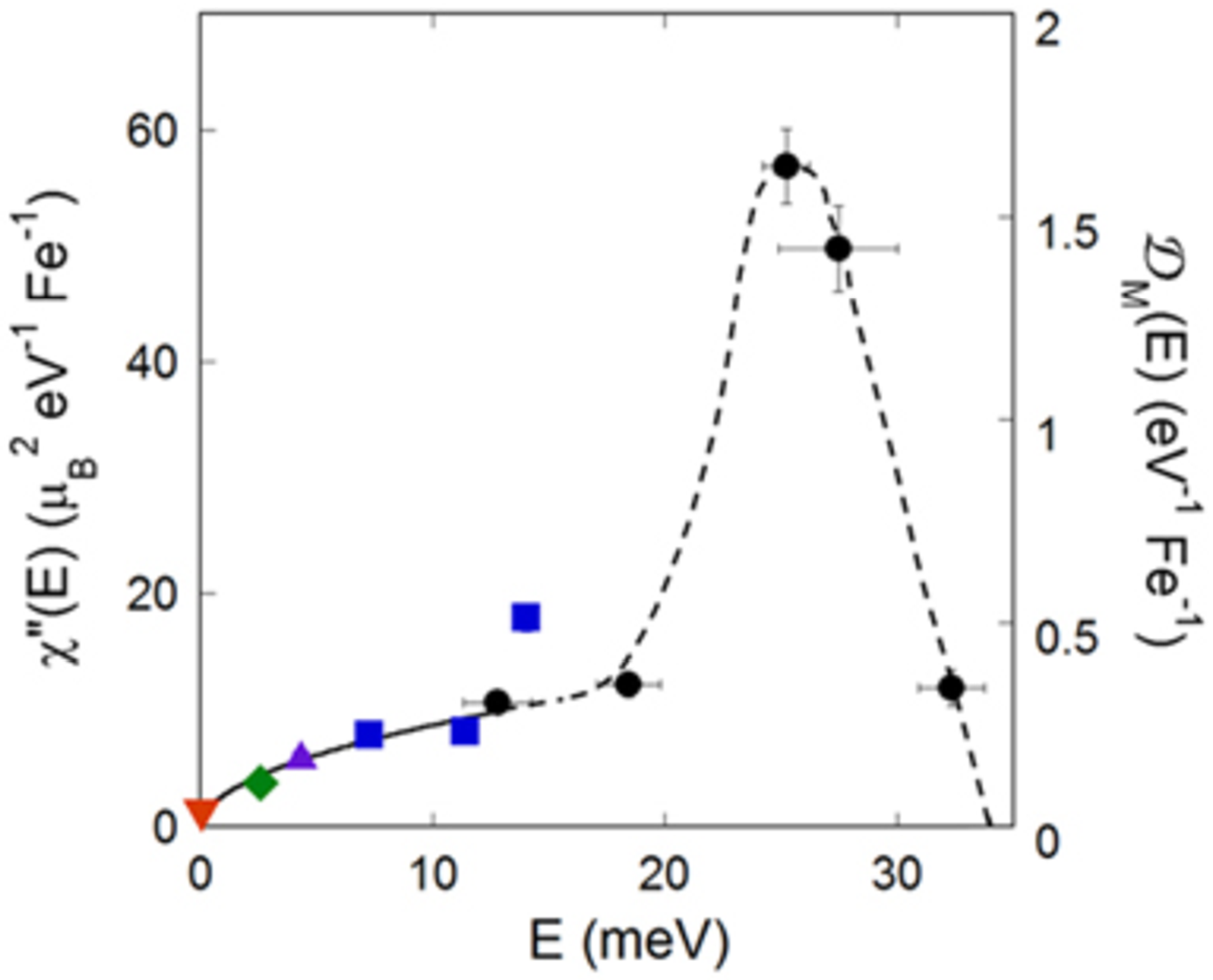}
\caption{(Color online) The $q$-integrated dynamical spin susceptibility $\chi''(E)$ for the lowest-$E$ magnon mode as a function of energy. The downward triangle was obtained at DNA. The diamond was obtained at AMATERAS. Upward triangle, squares, and circles were at 4SEASONS from $E_i$ = 12.5, 21.5, and 45.3 meV, respectively. The solid line is the fitting with a single parameter $\chi''_0$ using Eq. (4). The dashed line is a guide to the eye.}
\label{fig:9}
\end{figure}

The magnons of YIG extended up to 86 meV as shown in Fig. 6. Phonons were not apparently observed in this low-$Q$ region. The strong high-$E$ magnetic excitations were observed as nearly $Q$ independent dispersions at about 73 and 86 meV. In the middle-$E$ range below 55 meV, many dispersions overlap each other, leading to a broad band down to 30 meV. Three nearest-neighbor-exchange integrals, $J_{aa}$, $J_{ad}$, and $J_{dd}$, were estimated step by step based on the simulation with $gS$=5 $\mu_{\rm B}$ using {\sc spinw} software \cite{spinw} as follows. The subscripts $a$ and $d$ refer to the Fe 16$a$ (octahedral) and 24$d$ (tetrahedral) sites in the cubic symmetry  $Ia\bar{3}d$, respectively. 

$J_{ad}$ was determined from the whole magnon band-width, while $J_{dd}$ is determined by the magnon energy at $P$ ($\sim$45 meV) with positive chirality in the middle-$E$ range. A strong positive correlation was found between $J_{ad}$ and $J_{aa}$, which was sensitive to the second-highest-magnon energy at $P$ ($\sim$70 meV). The obtained three nearest-neighbor-exchange integrals, $J_{aa}$, $J_{ad}$, and $J_{dd}$, were 0.00$\pm$0.05, $-2.90$$\pm$0.07, and $-0.35$$\pm$0.08 meV, respectively. The minus sign means that the couplings are antiferromagnetic \cite{Barker}. The errors of integrals were determined using the largest energy shift up to 2 meV in the dispersion energies typically at the $P$ point. The calculated dispersions are shown in Fig. 6. Note that the present three exchange integrals are nearly consistent with estimated values ($J_{aa}$$\sim$0, $J_{ad}$=$-$2.78, $J_{dd}$=$-$0.28 meV) from magnetic susceptibility above 750 K\cite{Wojtowicz} after taking into account the temperature dependence of the lattice constant. As discussed in the Ref.\cite{Wojtowicz}, $J_{aa}$ is estimated to be less than $-$0.03 meV from the garnet compound Ca$_{3}$Fe$_{2}$Si$_{3}$O$_{12}$ with Fe$^{3+}$ occupied only at the 16$a$ site \cite{Geller}. In the previous analysis of magnon dispersions measured below 55 meV in Ref. \onlinecite{Plant}, $J_{aa}$, $J_{ad}$, and $J_{dd}$ were $-$0.69, $-$3.43, and $-$0.69 meV, respectively. After detailed refinement of the same dispersions, $J_{aa}$, $J_{ad}$, and $J_{dd}$ became $-0.33$, $-3.43$, and $-1.16$ meV, respectively\cite{Cherepanov}. The simulated magnon dispersions with these integrals seem consistent with those below 55 meV, but deviate largely from observed dispersions above 55 meV. The simulated magnon dispersions in ref.\onlinecite{Barker} also exhibit similar behavior.    
In order to check the validity of our exchange integrals, observed and simulated constant-$E$ cuts at various energies are shown in Fig. 7. They are fairly consistent with each other even in the middle-$E$ range from 30 to 50 meV, where many modes overlap with each other. The precise fitting may require more parameters than ours as discussed in recent studies \cite{Xie, Boothroyd}. 

As for the lowest-$E$ acoustic magnon dispersion, a quadratic dispersion is observed from data sets measured at various spectrometers below 14 meV near the $\Gamma$ point as shown in Fig. 8. The nearly isotropic low-$E$ dispersion can be written approximately as follows. 

\begin{equation}
E=Da^2q^2,
\label{eq:1}
\end{equation}
where $D$ is the stiffness constant, $q$ is the magnon wave vector, and $a$ is the lattice constant. $Da^2$ is estimated to be 633$\pm$17 meV\AA$^2$ (3.95 x 10$^{-29}$ erg cm$^2$=3.95 x 10$^{-40}$ J m$^2$) based on the fitting below 14 meV in Fig. 8. This value is slightly smaller than that of 670 meV\AA$^2$ (4.2 x 10$^{-29}$ erg cm$^2$) used in ref. \onlinecite{Kikkawa}. By using the obtained exchange integrals, the stiffness constant $D$ can be estimated as follows \cite{Srivastava, Cherepanov}.

\begin{equation}
D=\frac{5}{16}\left(8J_{aa}-5J_{ad}+3J_{dd}\right).
\label{eq:2}
\end{equation}

This equation leads to $Da^2$=642 meV\AA$^2$ from our three exchange integrals. This value is consistent with the stiffness constant obtained from Eq. (1). 

\subsubsection{Dynamical spin susceptibility}

The imaginary part of dynamical spin susceptibility $\chi''({\bf q}, E)$ is estimated based on the following equation for the magnetic differential scattering cross section:

\begin{widetext}
\begin{equation}
 \left(\frac{d^2 \sigma}{d \Omega dE}\right)_{M}= \frac{(\gamma r_{\rm e})^2}{\pi g^2 \mu_{\rm B}^2} \frac{{\bf k}_f}{{\bf k}_i} f^2(Q) t^2({\bf Q}) \{1+(\hat{\tau} \cdot \hat{\eta})^2\}_{av}\{1+n(E)\} \chi''({\bf q},E)\;,
 \label{eq:3}
\end{equation}
\end{widetext}
where the constant value  $(\gamma r_e)^2$=0.2905 barn sr$^{-1}$,  $g$ is the Land\'{e} $g$ factor; ${\bf k}_i$ and ${\bf k}_f$ are the incident and final wave vectors; the isotropic magnetic form factor $f^2(Q)$ of Fe$^{3+}$ at (220) is 0.8059 ($Q$=1.44 \AA$^{-1}$); the dynamic structure factor $ t^2({\bf Q})$\cite{Shirane} is approximated to be a squared static magnetic structure factor relative to full moments, i.e., $ t^2({\bf Q})$$\approx$$F^2_M(${\bf G}$)$/$F^2_{M0}$ =13/25; $\hat{\tau}$ is a unit vector in the direction of ${\bf Q}$; $\hat{\eta}$ is a unit vector in the mean direction of the spins; the angle-dependent term $\{1+(\hat{\tau} \cdot \hat{\eta})^2\}_{av}$ is 4/3 due to the domain average without a magnetic field; and $n(E)$ is the Bose factor.

The obtained imaginary part of $q$-integrated dynamical spin susceptibility $\chi''(E)$ is shown in Fig. 9. The $E$ dependence of $\chi''(E)$ for a quadratic dispersion case becomes a square-root function of energy\cite{Shirane}. In the case of a ferromagnet Fe, the constant-$E$ scan intensity of magnons with a certain $E$ width is inversely proportional to the slope of dispersion ($\propto$$1/\sqrt{E}$) because the integration in a $q$ range with a certain $E$ width for the quadratic dispersion is proportional to $1/\sqrt{E}$ \cite{Shirane}. Because of the magnon dispersion, the excitation at a finite energy $E$ appears at $q$ positions deviating from the $\Gamma$ point, forming a 3D $q$ spherical shell. For the intensity integration of the whole $q$-spherical-shell, the surface area $\sim4\pi q^2$ is proportional to the energy. The multiplication of $E$ by $1/\sqrt{E}$ results in $\sqrt{E}$ for the $q$-integrated intensity by a constant-$E$ scan. The same $E$-dependence of $\chi''(E)$ is expected in this ferrimagnet YIG at around the $\Gamma$ point because of the quadratic dispersion as follows. 
\begin{equation}
\chi''(E)=\chi''_0 \sqrt{E}.
\label{eq:4}
\end{equation}

Although the data are taken under five different conditions, all the values follow the same trend below 14 meV, which can be reproduced by Eq. (4) with a single parameter $\chi''_0$. The fitted value below 14 meV in Fig. 9 was 88$\pm$4  $\rm \mu_B^2$eV$^{-1.5}$Fe$^{-1}$. This nice fitting supports the validity of the theoretical model of LSSE\cite{Kikkawa} based on the MDOS estimated from the simple quadratic magnon dispersion only below 14 meV. Under this condition, the MDOS, ${\mathcal D}_M$ in our simple model can be described by the stiffness constant $D$ at $n(E)=0$. In addition, the MDOS is also proportional to the normalized $\chi''(E)$ obtained for the lowest-$E$ branch as follows.  
\begin{equation}
{\mathcal D}_M(E)= \frac{n_{mode}D^{-3/2}}{(2\pi)^2 40}\sqrt{E}= \frac{A\chi''_0}{g^2 \mu_{\rm B}^2 S(S+1)}\sqrt{E},
\label{eq:5}
\end{equation}
where $n_{mode}$ is a magnon mode number, $A$ is a constant value, and 40 is the number of Fe sites in the crystal unit cell with a cubic lattice parameter $a$=12.36 \AA. The value $g^2 \mu_{\rm B}^2 S(S+1)$ is 35 ${\rm \mu_B}^2$Fe$^{-1}$ for Fe$^{3+}$.
In the magnetic unit cell, however, there are only 20 Fe sites. Note that neither the site number nor the unit-cell volume changes Eq. (5) because the MDOS is proportional only to the volume per Fe site. In the calculation using the {\sc spinw} software \cite{spinw}, there are 20 modes in the first Brillouin zone for 20 Fe sites. Here, we focus on the lowest-$E$ acoustic branch with positive chirality.  

As for the constant value $A$, it is basically unity in a single-mode case because of the sum rule for $\chi''(E)$ in the $E$ integration at $n(E)=0$. Based on our experimentally obtained stiffness constant of 633 meV\AA$^2$, the constant value $A$ became 0.94$\pm$0.02 at $n_{mode}$=1 (single-mode case). Thus we confirmed the single mode for the lowest-$E$ magnon branch. Equation (5) can be regarded as a Debye model of magnons. The difference of the constant value from unity may be attributed to the approximation in our simple model in addition to our experimental errors. 

Above 14 meV, however, the magnon dispersion deviates from the quadratic function, resulting in the upturn of $\chi''(E)$ in Fig. 9. It is schematically shown as a dashed line in Fig. 9. For example, if the dispersion energy becomes proportional to the wave vector, $\chi''(E)$ increases quadratically. Then, it becomes zero at the highest-$E$ end of the mode. The estimation of $\chi''(E)$ in Fig. 9 is limited in the Brillouin zone for the cubic unit cell with 40 Fe ions. Therefore, there is a certain ambiguity for the value above 30 meV due to an overlap with a mode in another neighboring Brillouin zone. The validity of our estimation can be checked by the energy integration of $\chi''(E)$ for one mode in Fig. 9. It is roughly consistent with the theoretical value, 1/40 Fe$^{-1}$.  


Let us discuss the meaning of the single mode. We suggest that the mode is only a single chirality, as expected theoretically, although the present nonpolarized inelastic neutron scattering cannot distinguish two chiralities. This contrasts with doubly degenerate modes in the lowest-$E$ magnon dispersion of an antiferromagnet, which often split in $Q$ due to the Dzyaloshinskii-Moriya interaction \cite{Park}. 
The two types of chirality modes in YIG are split in energy due to the energy splitting of up and down spins. A polarized inelastic neutron scattering measurement would reveal the chirality of each magnon mode for a single-domain YIG crystal. This kind of experiment can be carried out under a magnetic field parallel to the scattering vector, which requires the estimation of a phase shift due to Larmor precession under the magnetic field. Anyway, it was proved here that there is only a single magnon mode at the lowest-$E$ branch in YIG, which has positive chirality based on our theoretical simulation.
\section{CONCLUSIONS} 

We have studied the basic parameters of YIG. The refined nuclear structure was distorted to a trigonal space group of $R\bar{3}$. As for the magnons, the highest-$E$ mode extended to 86 meV. Based on the whole magnon dispersions, the nearest-neighbor-exchange integrals, $J_{aa}$, $J_{ad}$, and $J_{dd}$, were estimated. The stiffness constant $D$ of a magnon dispersion below 14 meV was consistent with these estimated nearest-neighbor-exchange integrals. The imaginary part of the $q$-integrated dynamical spin susceptibility $\chi$"($E$) exhibited a square-root $E$ dependence in the energy range. Thus the applicable upper energy limit for the simple dispersion model of LSSE was about 14 meV. The lowest-$E$ magnon branch was found to have a single chirality mode based on our absolute-scale estimation of MDOS, which was consistent with a theoretical prediction\cite{Barker}. 

\begin{acknowledgments}

The work at J-PARC was performed under proposals 2012B0134 (BL01), 2013B0278 (BL14), 2015A0174 (BL01), 2014B0157 (BL02), 2015I0002 (BL02), 2016A0318 (BL02), and 2017A0222 (BL18). We acknowledge Prof. G. E. W. Bauer, Prof. Y. Nambu, Prof. K. Kakurai, Prof. E. Saitoh, Prof. S. Maekawa, Dr. R. Kajimoto, Dr. S. Toth, Dr. T. Kikkawa, Dr. Y. Ohnuma, and Dr. M. Mori for discussions, and the CROSS sample environment team for experimental assistance. This work was supported by JPSJ kAKENHI Grant No. JP25287094. H.O. is grateful for the use of supercomputers at the Japan Atomic Energy Agency and the Institute for Solid State Physics, the University of Tokyo. 
\end{acknowledgments}



\appendix

\section{Absolute intensity estimation}

Event-recording data sets obtained at neutron scattering spectrometers at the Materials and Life Science Experimental Facility of J-PARC have all the neutron detection time and position information for each detector pixel. Each of the $E_i$ data sets was extracted from the event-recording data using {\sc utsusemi} software\cite{Inamura}, and was converted to intensity data proportional to the scattering function, $I_{obs}({\bf Q},E)$$\propto$(${\bf k}_i$/${\bf k}_f$)$d^2$$\sigma$/$d\Omega dE$, as a function of $E$ for each detector pixel. $I_{obs}({\bf Q},E)$ was sliced for a specific region of ${\bf Q}$ and $E$ by using a slicer software {\sc viscontm} in {\sc utsusemi}. In this process, the intensity for each ${\bf Q}$ and $E$ bin gave an averaged intensity with a unit of counts sr$^{-1}$meV$^{-1}$r.l.u.$^{-3}$ of data points included in the bin. Note that the obtained intensity decreases when the specified region of the slice is expanded to a background region. This is due to averaging with the low-count background. On the other hand, the sliced intensity does not change by expanding the region to the no-detector region because there are no data points to be averaged in the expanded region. When one obtain an integrated intensity over a unit cell of the reciprocal lattice, its volume $\Delta V_{Q}$ (r.l.u.$^3$) depends on the reciprocal lattice specified in {\sc viscontm}. In our case, the original unit cell of YIG with $a$=12.36 \AA\  has 40 Fe sites (16$a$ and 24$d$), composed of 8 sub-unit cells with 5 Fe ions \cite{Plant}. In our crystal-setting, the measuring reciprocal zone was formed by three orthogonal reciprocal lattice vectors ($2,-1,-1$), ($1,1,1$), and ($0,-1,1$). In this crystal-setting Brillouin zone, the modified unit cell had 40/6 Fe ions (=$N_m$). In order to get the averaged dynamical spin susceptibility per Fe ion, the integration should be made in a reciprocal zone that effectively has only one Fe ion. 
So the $q$-integrated intensity (counts sr$^{-1}$meV$^{-1}$Fe$^{-1}$) was obtained from multiplying $I_{obs}(E)$(counts sr$^{-1}$meV$^{-1}$r.l.u.$^{-3}$) by the effective reciprocal space volume per Fe, $\Delta V_Q$/$N_m$ (r.l.u.${^3}$Fe$^{-1}$). 

Then we have to integrate all magnon intensities in one Fe Brillouin zone.  Instead of the integration of the whole zone, it is also possible to estimate the averaged dynamical spin susceptibility by correcting an integrated intensity at a specific reciprocal point based on the dynamic structure factor $ t^2({\bf Q})$\cite{Shirane} in Eq. (3). Here, $ t^2({\bf Q})$ was approximated to be the squared static magnetic structure factor ratio $F^2_M({\bf G})$/$F^2_{M0}$. The magnetic structure factor $F_M({\bf G})$ is written as.
\begin{equation}
F_M({\bf G})=\sum_{j=1}^{N_0} \sigma_{j} \exp(i{\bf G} \cdot r_j),
\label{eq:a1}
\end{equation}
where $\sigma_j$ is +1 or $-$1 for the $j^{th}$ spin in a magnet with a total number of spins $N_0$. $F_{M0}$ is the summation of +1 for all the spins, resulting in $N_0$. For example, the (220) magnetic Bragg peak intensity is reduced from the full magnetic intensity by the squared magnetic structure factor ratio  $t^2({\bf Q})$$\approx$$F^2_M({\bf G})$/$F^2_{M0}$=(2$^2$+3$^2$)/5$^2$=13/25 at (220) for YIG, where 2 and 3 are the numbers of Fe atoms at the 16$a$ and 24$d$ sites in the sub-unit cell of Fig. 1, respectively. Then, the average dynamical spin susceptibility for the whole Brillouin zone was estimated from the low-$E$ magnon intensity at (220). 

 The angle effect between the scattering unit vector $\hat{\tau}$ and the mean spin unit vector $\hat{\eta}$ was included as the angle-dependent term $\{1+(\hat{\tau} \cdot \hat{\eta})^2\}_{av}$=4/3 of the randomly oriented domain case in Eq. (3).     
   
The $q$-integrated scattering function $S(E)$ (mbarn sr$^{-1}$meV$^{-1}$Fe$^{-1}$) was obtained from the observed $q$-integrated neutron scattering intensity $I_{obs}(E)$ at ${\bf Q}$ for a domain average case as follows.
\begin{equation}
S(E)=\frac{3 C I_{obs}(E) \Delta V_Q}{4 N_{m} A^* f^2(Q) t^2({\bf Q})},
\label{eq:a2}
\end{equation}
where $C$ is the normalization factor, which was obtained from vanadium incoherent scattering \cite{Xu}, $\Delta V_Q$ is the reciprocal space volume in r.l.u.$^3$ specified in {\sc viscontm} for intensity integration, $N_m$ is the number of magnetic ions in the crystal-setting zone, and $A^*$ is the neutron absorption coefficient. 
  
The imaginary part of the $q$-integrated dynamical spin susceptibility $\chi''(E)$ ($\rm \mu_B^2$eV$^{-1}$Fe$^{-1}$) was obtained from the $q$-integrated scattering function $S(E)$ as follows.
\begin{equation}
\chi''(E)=\frac{\pi g^2}{(\gamma r_{\rm e})^2}\mu_{\rm B}^2 S(E) \{1-\exp(-E/k_{\rm B}T)\},
\label{eq:a3}
\end{equation}
where $(\pi g^2)/(\gamma r_{\rm e})^2$=43.26 sr barn$^{-1}$. The Debye-Waller factor term was neglected because of the measured low-$Q$ reflections. 
 
As demonstrated by the present measurement at 4SEASONS with the multi-$E_i$ option\cite{Nakamura}, many magnons in a wide $Q$-$E$ space were simultaneously observed by single scan without any crystal rotation. It took only 1 day. The present study is possible because of the powerful capability. On the other hand, the measuring region is a scattering curved-surface in a four-dimensional space $(Q_a, Q_b, Q_c, E)$. Although the energy had a strong correlation with reciprocal vectors as the scattering curved surface, the energy direction was approximated as a constant-$E$ slice because of the small $E$ width. The strong correlation was observed mainly along the $Q_a$ direction in our crystal setting. In this case, precise $q$ positions at an energy can be measured along the $q_b$ and $q_c$ directions, which are shown in Figs. 7 and 8. The validity of the constant-$E$ slice approximation can be found in the $E$ dependence of $\chi''(E)$ at low energies in Fig. 9. The $E$-dependence is the same as that observed in ferromagnetic Fe magnons\cite{Shirane}. In the integration of one 3D spherical-shell magnon excitation, the energy in Fig. 9 is an average value with a certain $E$ width due to the correlation mainly with $Q_a$ in our study.


{}

\end{document}